\documentclass[twocolumn,footinbib,showpacs,showkeys,superscriptaddress] {revtex4-1}
\usepackage{epsfig}
\usepackage{graphicx,graphics}
\usepackage{epstopdf}
\usepackage{natbib}
\usepackage{lipsum}
\usepackage{dcolumn}
\usepackage{amsmath,amssymb,amsfonts}
\usepackage{latexsym,verbatim}
\usepackage{bm}
\usepackage{color}
\usepackage{siunitx}
\usepackage[breaklinks=true,colorlinks,citecolor=blue,linkcolor=blue,urlcolor=blue]{hyperref}

\begin{document}
\preprint{AIP/123-QED}	
\title{Modulation of energy and angular momentum radiation of two-dimensional altermagnets}
\author{Yong-Mei Zhang}
 \homepage{zymzym@nuaa.edu.cn}
\affiliation{College of Physics, Nanjing University of Aeronautics and Astronautics, Jiangsu 210016, People's Republic of China}
\author{Zhi- Ping Niu}
\affiliation{College of Physics, Nanjing University of Aeronautics and Astronautics, Jiangsu 210016, People's Republic of China}

\date{\today}

\begin{abstract}
This paper investigates the energy and angular momentum radiation of altermagnets under Rashba spin-orbit coupling (RSOC) and external magnetic fields. Using an effective low-energy Hamiltonian, we derive electronic energy bands and calculate optical conductivity via the Kubo formula. Results show that RSOC strength, altermagnet interactions strength , and N\'{e}el vector direction notably affect the optical conductivity of altermagnet metals. Energy radiation is highly sensitive to Rashba spin-orbit coupling, with a saturation effect beyond a particular value, and its peak emission rate is lower than that of graphene due to reduced conductivity. Different from usual semi-conductor, semi-metal or Dirac materials, eg. Graphene or silicene, altermagnets generate angular momentum radiation with specific Rashba spin-orbit coupling and altermagnet interaction strength. Angular momentum radiation is minimally reactive to  Rashba spin-orbit coupling at low altermagnet interactions strength  values but exhibits drastic oscillations between extreme values as  altermagnet interactions strength reaches a critical point, showing high sensitivity. These findings suggest that adjusting these parameters can tailor altermagnet applications in spintronics and quantum technologies, potentially leading to innovative devices with customized radiation attributes.
\end{abstract}
\maketitle

\section { INTRODUCTION }
Altermagnets, as a novel class of magnetic materials, are characterized by their unique properties that include zero net macroscopic magnetization and the breaking of time-reversal symmetry without the need for an external magnetic field, which is a key characteristic that distinguishes them from conventional ferromagnets and antiferromagnets \cite{ref17,ref18}. This time-reversal symmetry breaking (TRSB) is a result of their distinct crystal symmetries that inhibit magnetization while promoting strong spin polarization in the band structure. Studies have shown that altermagnets can generate spin currents transverse to the injection direction, with polarization along the N\'{e}el vector, through a process known as the spin splitting effect \cite{ref19,ref20}. This indicates that altermagnets have the ability to transfer angular momentum, as spin currents are a form of angular momentum transfer. Altermagnets are characterized by their collinear compensation magnetization, which results in spontaneous spin splitting in the electronic bands in the limit of zero spin-orbit coupling (SOC) \cite{ref21,ref22}. The N\'{e}el vector's direction is intrinsically linked to this spin splitting, and its strength and orientation are determined by the material's crystal symmetry. In some cases, the N\'{e}el vector's direction can connect sub-lattices with opposite spins not through translation or inversion, but by a spatial symmetry that involves a rotation or reflection.

All matter above absolute zero temperature radiates energy in the form of photons due to thermal motion. This is known as thermal radiation, which occurs because atoms and molecules within the material undergo energy level transitions due to their thermal motion, thereby emitting electromagnetic waves \cite{ref1,ref2,ref3}. In some cases, the radiation process is accompanied by the emission of angular momentum. For example, when charged particles move in a magnetic field, they emit radiation that carries angular momentum \cite{ref4,ref5}. This type of radiation is often referred to as vortex light or vortex radiation, characterized by a specific distribution of angular momentum. This phenomenon has been extensively studied in high-energy physics experiments, such as in free-electron lasers and undulators \cite{ref6,ref7}. Angular momentum radiation has significant applications across various fields, particularly in physics and communication technology \cite{ref8,ref9}.

Angular momentum radiation, especially orbital angular momentum (OAM), is of great significance in the field of communication. OAM beams have a helical phase front and a field singularity at the axis, which allows them to be used for information transmission, imaging, and particle manipulation \cite{ref10}. The number of orthogonal modes in OAM beams is theoretically infinite, and each mode can serve as an element of a complete orthogonal basis for multiplexing different signals, thereby significantly increasing spectral efficiency. The concept of OAM is not limited to electromagnetic waves; acoustic waves can also carry OAM \cite{ref11}. This opens new research areas for wireless communication and acoustic communication, especially underwater communication, and may break through the limitations of existing communication systems. Angular momentum radiation also has potential applications in the field of quantum optics and quantum information, such as quantum entanglement and quantum communication \cite{ref12,ref13}.
The generation of angular momentum radiation indeed requires time-reversal symmetry breaking  \cite{ref14,ref15}. This is because time-reversal symmetry implies that the system's behavior should be identical when the direction of time is reversed, which would not allow for a net transfer of angular momentum through radiation. Hu et al  \cite{ref16} discusses how an unconventional anisotropic spin-momentum interaction can lead to time-reversal symmetry breaking, which is decisive for the generation of angular momentum radiation. Based on the evidence from altermagnet research results, it is reasonable to infer that altermagnets do have the capacity to emit angular momentum. This is supported by their ability to generate spin currents, their response to optical excitation, and their unconventional spin-polarized band structures, all of which are indicative of angular momentum transfer and radiation. In altermagnets, the direction and strength of the N\'{e}el vector play a crucial role in determining the characteristics of spin splitting \cite{ref23,ref24}. The N\'{e}el vector's orientation can influence the spin polarization of the bands, which in turn affects the angular momentum radiation emitted by the material. The ability to manipulate the N\'{e}el vector through external fields or structural changes allows for control over the spin currents and the resulting angular momentum radiation \cite{ref13,ref25}. The crystallographic orientation of altermagnetic materials plays a critical role in their magnetic properties, electrical transport characteristics, and the realization of novel physical phenomena. By manipulating the crystallographic orientation, new ideas and methods can be provided for the development of high-performance magnetic materials and novel electronic devices.
In this paper, we study the angular momentum radiation of altermagnet material. We aim at investigating how the unusual properties of altermagnet materials modify angular momentum radiation by adjusting altermagnet strength and direction of N\'{e}el vector. The angular momentum radiation of altermagnet materials is influenced by their unique spin polarization and band structure. By adjusting the topological structure, spin-momentum locking, and domain wall dynamics, it is possible to effectively control and manipulate the angular momentum radiation characteristics of these materials. These research findings provide a theoretical foundation and experimental guidance for further exploration and development of altermagnet materials in the fields of spintronics and quantum information processing.
The structure of the paper is outlined as follows: In Section \uppercase\expandafter{\romannumeral2}, we introduce the model Hamiltonian of altermagnet materials and the procedure to calculate energy band and optical conductivity. Section \uppercase\expandafter{\romannumeral3} delves into the numerical results of radiation properties, with a particular focus on examining how they are affected by the strength of the Rashba spin-orbit coupling (RSOC) and N\'{e}el direction, as well as the influence of an external magnetic field. Section \uppercase\expandafter{\romannumeral4} provides a summary of the findings presented in the paper.

\section { MODEL AND METHOD}
\subsection{ Hamiltonian model}
We consider a $d$-wave altermagnet thin films with substrate induced Rashba spin-orbit coupling. Rutiles ($RuO_{2}$) are the prototypical representatives of d-wave altermagnetism \cite{ref26}. A N\'{e}el vector of arbitrary direction is determined by the intrinsic anisotropy and applied magnetic field \cite{ref27,ref28}. The total Hamiltonian of the system can be written as
\begin{equation}
	\label{HT}
	 H = \frac{p^2}{2m} + \alpha (\vec{\sigma} \times \vec{p}) \cdot \vec{e}_z - \mu_B \vec{\sigma} \cdot \vec{B} + H_{alt} .
\end{equation}

The first term signifies electron kinetic energy with $m$  the effective mass. The second term quantifies the Rashba spin-orbit coupling induced by the substrate with $\alpha$  the coupling strength. The third term describes Zeeman effect in the altermagnet material due to external magnetic field $\vec{B}$ . $\vec{\sigma} ( \sigma_{x},\sigma_{y},\sigma_{z})$  is the vector of spin Pauli matrices. The altermagnet term in Eq. \eqref{HT} due to the  $d$-wave metallic background is written as

\begin{equation}
	\label{eq_H_alt}
	H_{alt} = \frac{\beta}{2} (p_{x}^{2} - p_{y}^{2}) \hat{n} \cdot \vec{\sigma} .
\end{equation}

In which $\beta$  characterizes the strength of altermagnet interaction. The unit vector $\hat{n}$  is parallel to the background N\'{e}el magnetization.
In spherical coordinates, the N\'{e}el vector is expressed by polar angle ($\theta$) and azimuthal angle ($\varphi$ ) as $\hat{n}=(\sin{\theta} \cos{\varphi} ,\sin{\theta} \sin{\varphi}, \cos{\theta})^{T}$ . The polar angle $\theta$  is the angle between the positive $z$-axis and the radial vector from the origin to the point in space. The azimuthal angle $\varphi$  is the angle in the $xy$-plane from the positive  $x$-axis to the projection of the radial vector onto the $xy$ -plane. The total Hamiltonian can be explicitly written as
\begin{equation}
	\label{HT2}
\begin{split}
  H = & \frac{p^2}{2m} \sigma_0 + \left( \frac{\beta}{2} (p_x^2 - p_y^2) \sin\theta \cos\varphi + \alpha p_y \right) \sigma_x \\
    & + \left( \frac{\beta}{2} (p_x^2 - p_y^2) \sin\theta \sin\varphi - \alpha p_x \right) \sigma_y \\
    & + \left( \frac{\beta}{2} (p_x^2 - p_y^2) \cos\theta - \mu_B B \right) \sigma_z
\end{split}
\end{equation}

\begin{figure}
	\includegraphics[width=0.9\columnwidth]{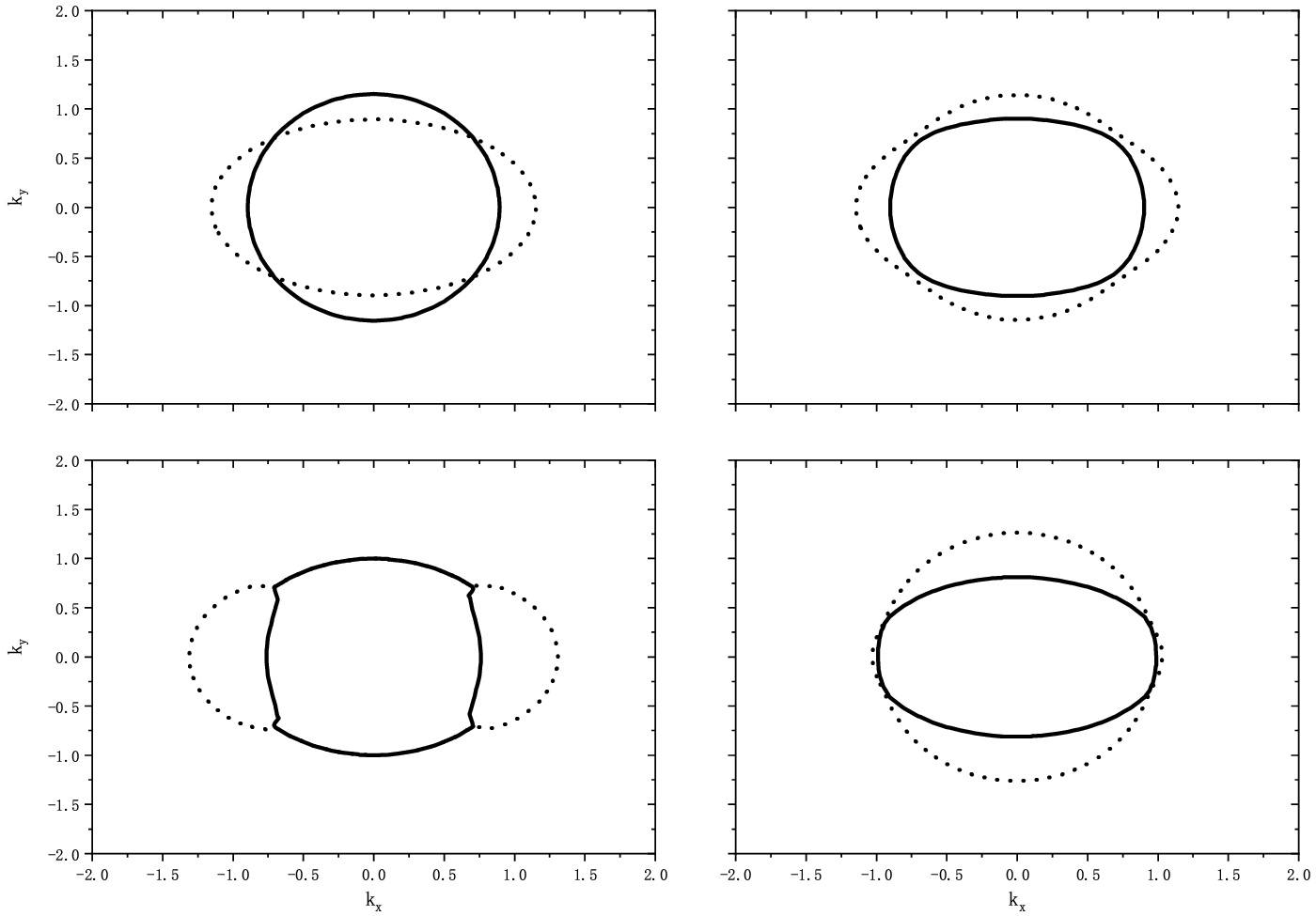}
	\caption{ Energy contour of electron band at $E=0.5eV$ . (a)$\beta=0.5$  ; (b)$\beta=0.5$ ,$\alpha=0.1$  ;(c) $\beta=0.5$  , $\theta=0.5\pi$ , $B=0.2$;  (d) $\beta=0.5$  , $B=0.2$ . Parameters that are not specified are zeroed. }
	\label{fig1}
\end{figure}

Here $\sigma_{0}$  is a 2-by-2 identity matrix. The eigenvalues of the Hamiltonian can be solved by diagonalizing the Hamiltonian in Eq.\eqref{HT2}. The main features of the electronic band are shown in  Fig. \ref{fig1}, for different sets of parameters.

In Fig. \ref{fig1} we plot energy contour of altermagnets energy band. Without spin-orbit coupling and magnetic field, the Fermi surface at a positive energy consists of two ellipses with spins polarized along $\pm \hat{n}$ . The two ellipses intersect at diagonals $p_{y}=\pm p_{x}$ , see Fig. \ref{fig1}(a). Spin-orbit coupling couples spin to momentum and splits elliptical Fermi surfaces at four intersections where $z$-components of spin vanish, see Fig. \ref{fig1} (b). When magnetic field is applied and N\'{e}el vector direction is along  $x$ axis, the Fermi surface of $E_{+}$  is distorted along $k_{y}$ , while  $E_{-}$ is elongated along $k_{x}$  direction, as seen in Fig. \ref{fig1}(c). Fig. \ref{fig1} (d). is the energy contour when the direction of N\'{e}el vector is along  $z$ axis.

In altermagnet materials, SOC can significantly affect the electronic band structure and, consequently, the optical properties. The strength $\beta$  of the altermagnet material causes the spin polarization and band splitting. The direction of altermagnet interaction is also essential as it determines the orientation of the spin polarization and the resulting magnetic anisotropy \cite{ref29}. The altermagnet strength and direction in the Hamiltonian of the system can alter the optical conductivity and the radiation spectrum.

\subsection{ OPTICAL CONDUCTIVITY AND RADIATION FORMULA}
When discussing the optical conductivity and radiation spectrum of altermagnet materials, several key factors come into play, including RSOC (denoted as $\alpha$ ), altermagnet interaction strength (denoted as $\beta$), and the direction of N\'{e}el vector (denoted as $\hat{n}$ ) as well as magnetic field $\vec{B}$ \cite{ref30} .
The optical conductivity of a material describes how it responds to electromagnetic radiation, particularly in the visible and infrared regions of the spectrum. It is a complex quantity that includes both the real part (which is related to the absorption of light) and the imaginary part (which is related to the scattering of light).
The optical conductivity
$\sigma(\omega) $  can be expressed in terms of the Kubo formula \cite{ref31,ref32,ref33} , which connects the current-current correlation function to the frequency-dependent conductivity:

\begin{equation}
	\label{eq-Sigma}
    \begin{aligned}
    \sigma_{\alpha\beta}(\omega)&=\frac{i e^2\hbar }{S} \sum_{m,n,{\bf{k}}} \frac{\langle n,{\bf{k}}|V^{\alpha}({\bf{k}})|m,{\bf{k}}\rangle \langle m,{\bf{k}}|V^{\beta}({\bf{k}})|n,{\bf{k}}\rangle}{E_m ({\bf{k}})-E_n ({\bf{k}})-(\hbar\omega + i\eta)} \\
    &\cdot\frac{f (E_m ({\bf{k}}))-f (E_n ({\bf{k}}))}{E_m ({\bf{k}})-E_n ({\bf{k}})}.
    \end{aligned}
\end{equation}

Here, $\alpha(\beta)$  is coordinates  $x$ or $y$ .
$ V^{\alpha(\beta)} = \frac{1}{\hbar} \frac{\partial H}{\partial k_{\alpha(\beta)}}$
is velocity matrix and can be derived from \eqref{HT2} .

\begin{equation}
\label{velocity_vx}
 V^x =
\begin{pmatrix}
&\frac{\hbar k_x}{m} + \beta \hbar k_x \cos\theta
&\beta \hbar k_x \sin\theta e^{-i\varphi}+i\hbar\alpha   \\
&\beta \hbar k_x \sin\theta e^{i\varphi} -i\hbar\alpha
&\frac{\hbar k_x}{m} - \beta \hbar k_x \cos\theta
\end{pmatrix}
\end{equation}

\begin{equation}
\label{velocity_vy}
 V^y =
\begin{pmatrix}
&\frac{\hbar k_y}{m} - \beta \hbar k_y \cos\theta
&-\beta \hbar k_y \sin\theta e^{-i\varphi}+\hbar\alpha   \\
&-\beta \hbar k_y \sin\theta e^{i\varphi}+\hbar\alpha
&\frac{\hbar k_y}{m} + \beta \hbar k_y \cos\theta
\end{pmatrix}
\end{equation}

The optical conductivity characteristics of a material significantly affect its thermal radiation properties by influencing the carrier concentration, thermal conductivity, thermal effects, bandgap width, and dynamic regulation capabilities. The radiation spectrum refers to the distribution of radiation intensity over different wavelengths or frequencies \cite{ref34,ref35}
\begin{equation}
\label{Sw}
\left\langle S_{z}(\omega) \right\rangle = \frac{1}{6\pi^{2}\varepsilon_{0}c^{3}} \hbar \omega^{3} N(\omega) \operatorname{Re}\left[\sigma_{xx}(\omega) + \sigma_{yy}(\omega)\right]
\end{equation}

\begin{equation}
\label{Nw}
\left\langle N_{z}(\omega) \right\rangle = \frac{1}{12\pi^{2}\varepsilon_{0}c^{3}} \hbar \omega^{2} N(\omega) \operatorname{Im}\left[\sigma_{xy}(\omega) - \sigma_{yx}(\omega)\right]
\end{equation}

From the above formula we know that the optical conductivity and radiation spectrum of altermagnet materials are determined by a complex interplay of factors, including spin-orbit coupling, altermagnet strength and direction, and the presence of an external magnetic field. These factors must be considered when modeling and analyzing the optical properties of these materials. The total energy and angular momentum radiation are obtained by integrating radiation spectrum over the entire frequency range.

\begin{equation}
\label{Sz}
\left\langle S_{z} \right\rangle = \int_{0}^{\infty} \left\langle S_{z}(\omega) \right\rangle \, d\omega
\end{equation}

\begin{equation}
\label{Nz}
\left\langle N_{z} \right\rangle = \int_{0}^{\infty} \left\langle N_{z}(\omega) \right\rangle \, d\omega
\end{equation}

\section { RESULTS AND DISCUSSION}
\subsection{Optical conductivity}
Based on the above formula, optical conductivity of altermagnet material is calculated. It is observed from Eq. \eqref{Sw} and \eqref{Nw} that only real part of longitudinal conductivity and imaginary part of transverse conductivity contribute. In the following we'll pay more attention to these components.

\begin{figure}
	\includegraphics[width=0.9\columnwidth]{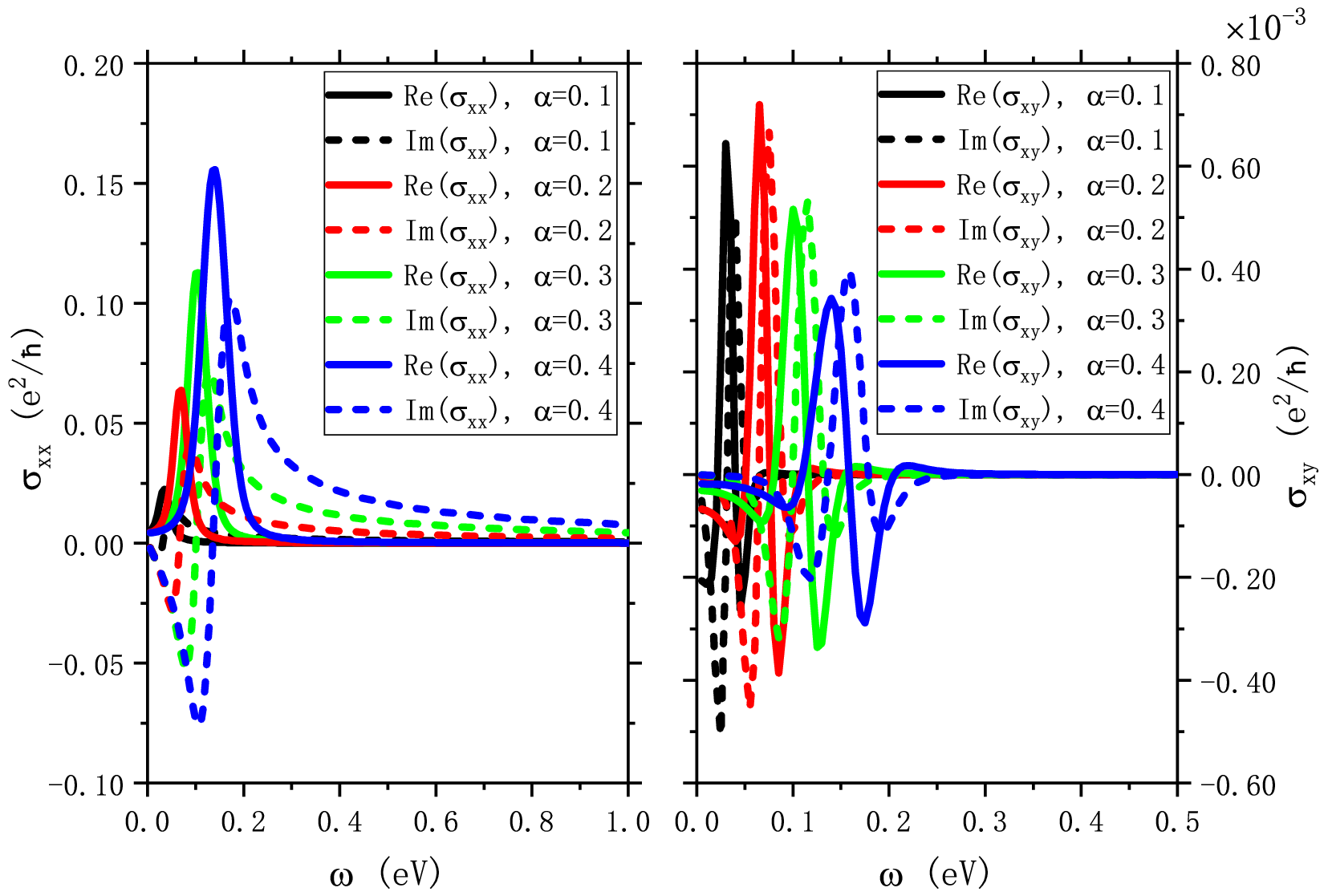}
	\caption{ Optical conductivity of altermagnet with different spin-orbit coupling strength. The curves in different color correspond to different RSOC strength. Solid curves for real part and dashed curves for imaginary part. Other parameters are set $\theta=0.25\pi$, $\varphi=0.00$, $B=0.00$, $\mu=0.20$, $\beta=0.40$, $T=300K$. }
	\label{fig2}
\end{figure}

From Fig. \ref{fig2} it is observed the role of RSOC strength $\alpha$  played on longitudinal and transverse conductivity. The curves in different color correspond to different RSOC strength. Solid curves represent real part and dashed curves for imaginary part. The real part of the longitudinal conductivity typically exhibits a Lorentzian line shape, which is a result of state lifetime broadening due to processes like collisions or spontaneous emission. For small RSOC strength, the peak is low and centered at low frequencies. As  RSOC strength   increases, the peak becomes higher and shifts to higher frequencies. The imaginary part of the longitudinal conductivity varies from negative to positive with frequency, which resembles Fano line shape.

As for the transverse conductivity, both the realc and imaginary parts exhibit complex oscillations, with positive and negative peaks. The overall value is quite small compared to the longitudinal component. Fano line shapes are due to interference between a direct path and a resonant auxiliary indirect path, which can produce extremely sharp line shapes and sudden changes over a very narrow wavelength range. These characteristics are fundamental for understanding the electromagnetic response of altermagnetic materials and for designing high-performance optical devices. This is quite different from diamagnetism like graphene. It is even different from silicene with Rashba spin-orbit coupling. Both graphene and silicene have zero transverse conductivitgy.

\begin{figure}
	\includegraphics[width=0.9\columnwidth]{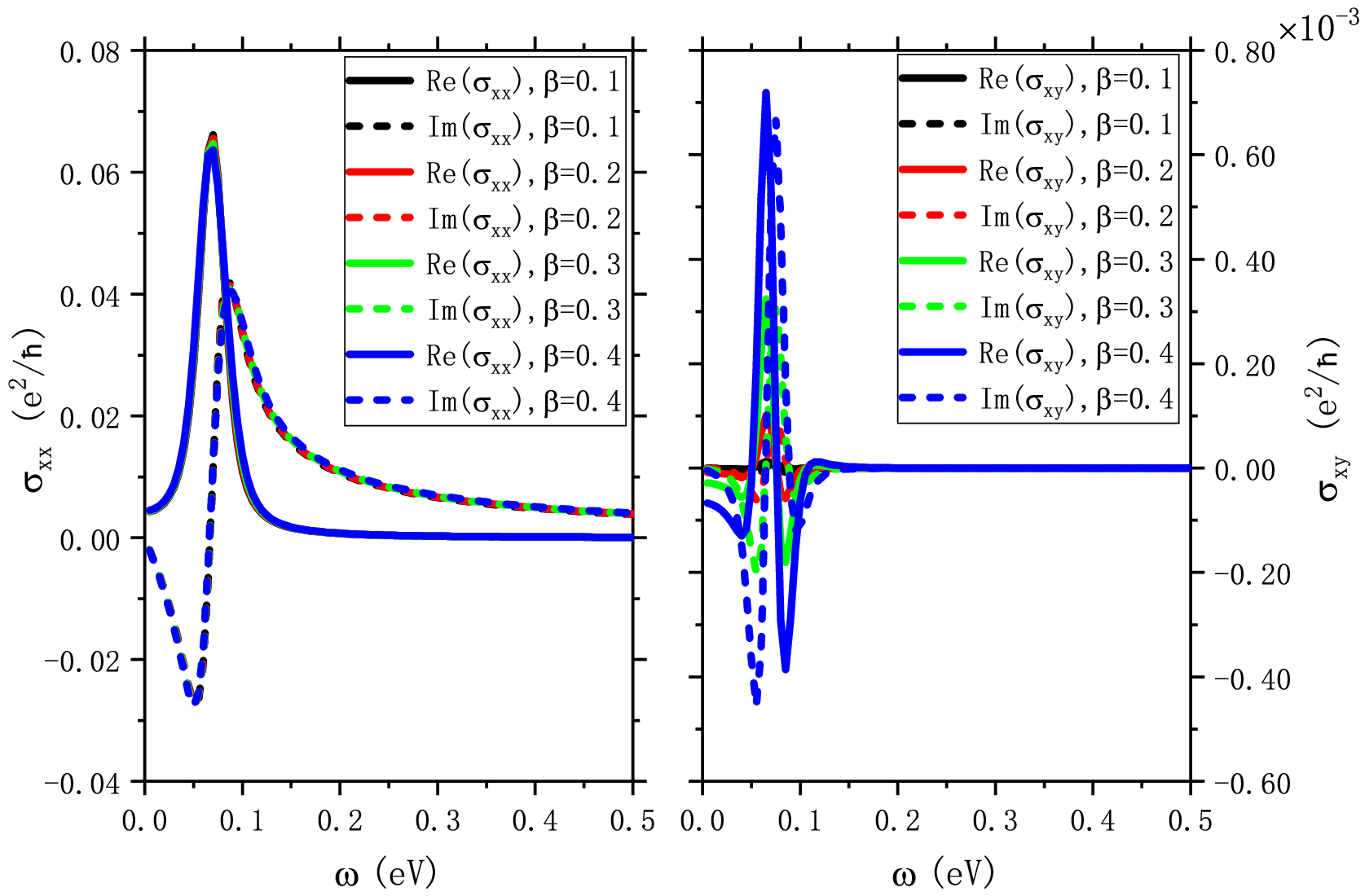}
	\caption{ Optical conductivity of altermagnet with different altermagnet strength. $\alpha=0.1$, $\theta=0.25\pi$, $\varphi=0.00$, $B=0.00$, $\mu=0.20$, $T=300K$. }
	\label{fig3}
\end{figure}

 We pay more attention on the effect of altermagnet interaction $\beta$  on optical conductivity since parameter $\beta$ describes main property of altermagnets.  As seen in Fig.\ref{fig3}, longitudinal conductivity exhibits narrow peak at low frequency, which is quite different antiferromagnetic materials \cite{cheng2025thermal} or ferromagnetic material \cite{ref34}. Peak value is quite small compared with graphene or semid-Dirac materials. It changes slightly with variations in altermagnet interaction strength. This reason might be that longitudinal conductivity is more influenced by the internal electron density and electron mobility of the material, which are not directly related to altermagnet interaction. The right panel of Fig.\ref{fig3} shows transverse conductivity varies noticeably with altermagnet interaction. Transverse conductivity is mainly located at low frequencies with thin peaks. Peak value increases with the increasing of altermagnet interaction. This could be ascribed to the fact that transverse conductivity is influenced by the anisotropy within the material's structure, and  altermagnet interaction is responsible for this anisotropy. Despite the noticeable change, the actual impact might be limited because the absolute value of transverse conductivity is small.
 In summary, altermagnet interaction has little effect on longitudinal conductivity, while it does have an evident effect on transverse conductivity, albeit a small one. This implies altermagnet interaction strength is an ideal tool to modify angular momentum radiation. Local magnetic in altermagnet makes it resemble properties of ferromagnet in the aspect of transverse conductivity. The optical conductivity properties are inevitably reflected in the radiation characteristics.

\begin{figure}
	\includegraphics[width=0.9\columnwidth]{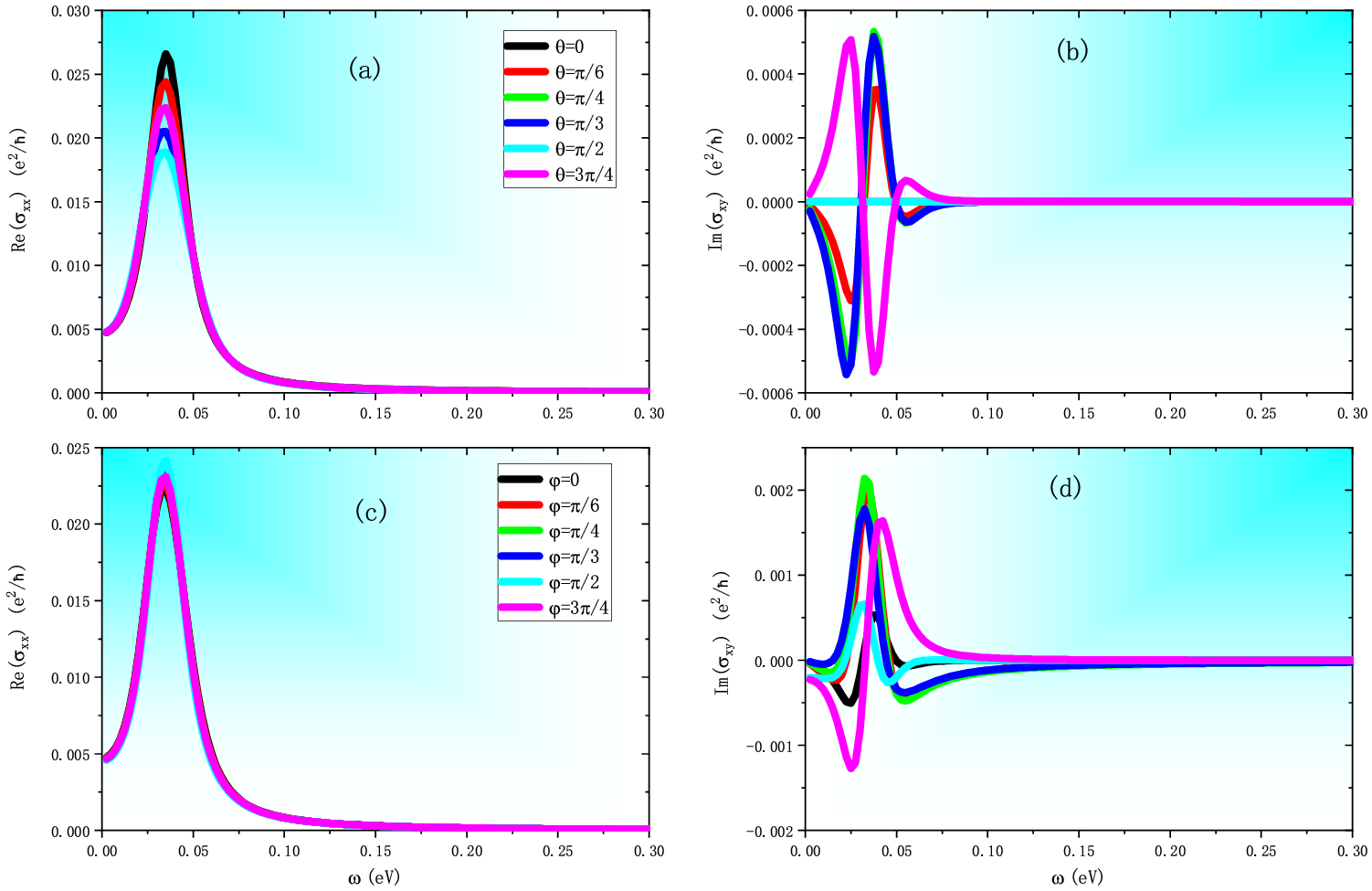}

	\caption{ Optical conductivity with various N\'{e}el vector directions. Parameters are set $\alpha=0.1$ ,$B=0.00$, $\mu=0.20$, $\beta=0.40$, $T=300K$. (a) and (b) show $Re(\sigma)_{xx}$  and $Im(\sigma)_{xy}$  change with polar angle $\theta$  with $\varphi=0.0$ . (c) and (d) Optical conductivity changes with azimuthal angle $\varphi$  with $\theta=0.25\pi$ . }
	\label{fig4}
\end{figure}

The direction of N\'{e}el vector is determind by the polar angle $\theta$ and the azimuthal angle $\varphi$. In Fig.4 the effect of N\'{e}el vector on optical conductivity is demonstrated. Optical conductivity is pronounced only at very low frequencies. Both the polar angle $\theta$  and the azimuthal angle $\varphi$  have minimal effects on the real part of longitudinal conductivity, e.g. $Re(\sigma)_{xx}$ . Regardless of how the angles change, the peak positions of longitudinal conductivity remain stationary. However, longitudinal conductivity decreases slightly as $\theta$ increases and increases slightly as  $\varphi$  increases.

Imaginary part of transverse conductivity $Im(\sigma)_{xy}$  changes strikingly with frequency $\omega$  from negative to positive when  $\theta$  or $\varphi$  varies within the range of zero to $\pi/2$ . As  $\theta$   increases within this range, the absolute value of transverse conductivity increases. When $\theta$  equals zero or  $\pi/2$ , the transverse conductivity vanishes .  Transverse conductivity  at $\theta=\frac{3}{4}\pi$  is opposite to that at $\theta=\frac{1}{4}\pi$  , showing antisymmetric about the $xy$ plane . When $\varphi$  increases, transverse conductivity  first increases and then decreases. The maximum value occurs when $\varphi=\pi/4$  . As  $\varphi$ passes $\pi/2$  , the dip and peak positions of $Im(\sigma)_{xy}$   shift to opposite. These numerical results imply that polar angle  $\theta$   and the azimuthal angle $\varphi$  have a pronounced effect on the imaginary of transverse conductivity. In other words, the direction of N\'{e}el vector is more crucial to angular momentum radiation.

\begin{figure}
	\includegraphics[width=0.9\columnwidth]{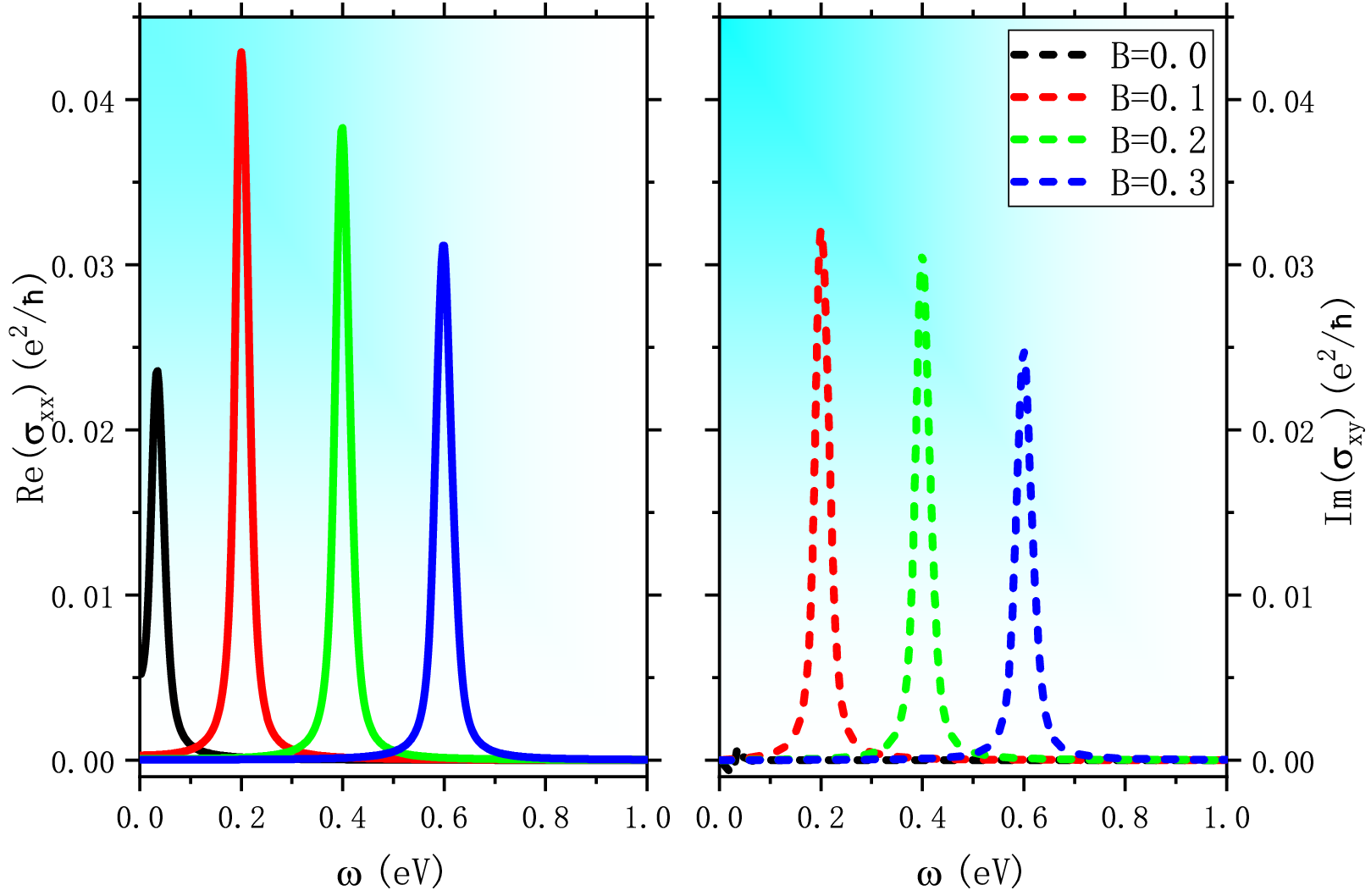}
	\caption{  Optical conductivity with variation of magnetic field. Parameters are set $\theta=\pi/4$,$\varphi=0.0$,$\mu=0.2$,$\alpha=0.1$,$\beta=0.5$,$T=300K$. }
	\label{fig5}
\end{figure}

Upon the application of a magnetic field, both the longitudinal and transverse conductivities exhibit a pronounced peak at a frequency $\omega$, see Fig.5. With an increase in the strength of the magnetic field, the conductivity peak is observed to migrate towards higher frequencies. This observation underscores the substantial influence that a magnetic field exerts on both longitudinal and transverse conductivities. Moreover, it suggests that the magnetic field serves as an effective measurement to modulate the radiation spectrum to specific photon frequency or high energy range.
Under the combined influence of various parameters, the electrical conductivity of altermagnets exhibits a relatively narrow peak at lower frequencies. This is in stark contrast to the broad frequency-domain conductivity of materials like graphene. Consequently, this will also affect the radiation properties of altermagnets, making them significantly different from those of Dirac materials, and even semi-Dirac materials.

\subsection{ RADIATION SPECTRUM}

\begin{figure}
	\includegraphics[width=0.9\columnwidth]{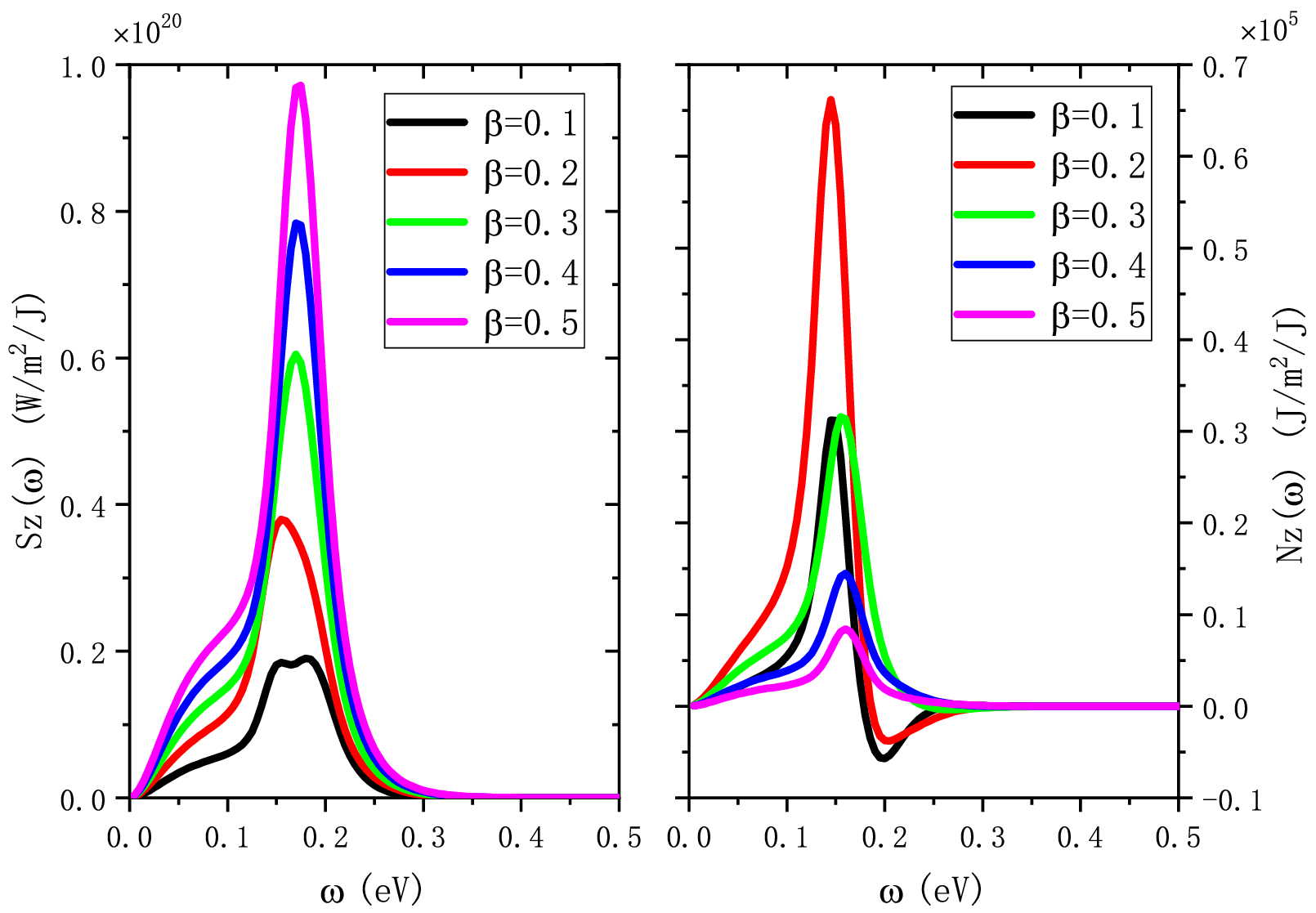}
	\caption{ Energy and angular momentum radiation spectrum of altermagnet with different altermagnet strength. Other parameters are listed in the graph. }
	\label{fig6}
\end{figure}

Although the strength of altermagnet interaction does not affect optical conductivity much, it has pronounced effect on radiation spectrum, see Fig.6. There is a narrow peak in energy radiation spectrum centered around $0.174 eV$($\sim 4.05\mu m$ ). The position of this peak does not shift when  $\beta$  increases. However, the peak value increases significantly, and the slope becomes steeper. Angular momentum radiation oscillates with frequency and transitions from positive to negative values when altermagnet interaction is small. Photons of different frequency have opposite angular momentum radiation. As altermagnet interaction increases, angular momentum radiation decreases gradually and remains always positive. Altermagnet interaction generates angular momentum radiation, similar to the effect of magnetic field effect in ferromagnets. However, angular momentum radiation from altermagnet materials is much weak compared with that from magnetic system due to quite weak local magnetic characteristic. Both energy radiation and angular momentum radiation of altermagnet are similar to that of ferromagnetic system. Energy radiation of altermagnet is 2 orders greater than that of antiferromagnetic materials. Altermagnet is in contrast different with antiferromagnet in angular momentum radiation Angular momentum radiation from altermagnet is the same order with ferromagnet, while there is no angular momentum radiation generated from antiferromagnetic materials.

\begin{figure}
	\includegraphics[width=0.9\columnwidth]{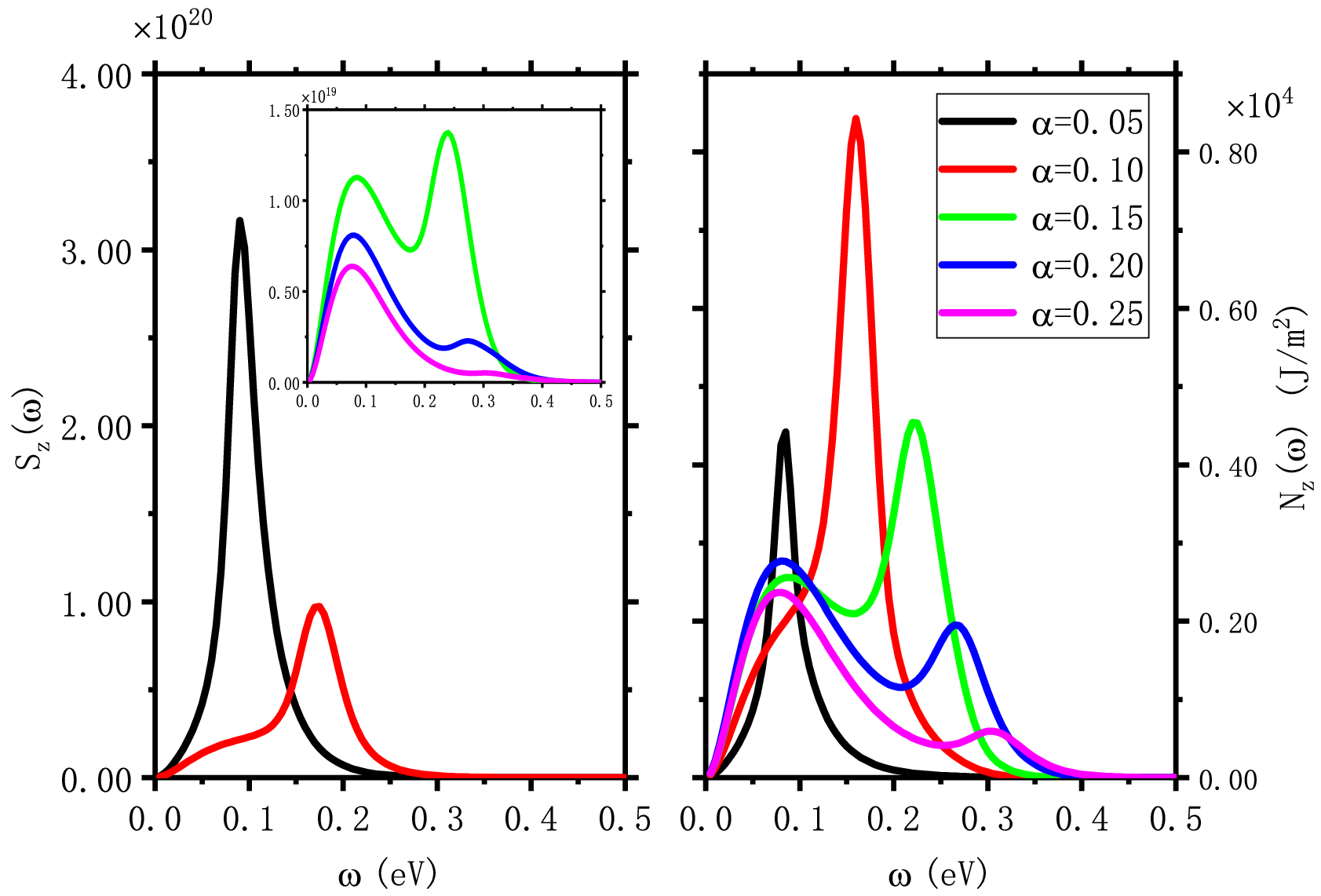}
	\caption{ Radiation spectrum of altermagnet with different spin-orbit coupling strength. Other parameters are set the same as Fig.\ref{fig3}. }
	\label{fig7}
\end{figure}

The radiation power exhibits considerable variation in response to the spin-orbit coupling. When the strength of the Rashba spin-orbit coupling (RSOC) $\alpha$ is relatively small, the radiation power is elevated and predominantly centered around a lower frequency $0.09eV$ , which is $~7.83\mu m$ in wavelength. However, as $\alpha$  escalates, the radiation power not only diminishes significantly but also progressively bifurcates into two distinct peaks. These peaks increasingly diverge from one another, with the principal peak migrating between them. As for the angular momentum radiation spectrum, a single peak is observed when the strength of the spin-orbit coupling is small. As Rashba spin-orbit coupling strength increases, the radiation spectrum tends to bifurcate, resulting in two peaks. The left peak remains nearly stationary in its position, whereas the right peak shifts towards a higher frequency and experiences a gradual decrease in intensity. This is because, as the strength of the Rashba Spin-Orbit Coupling (RSOC) increases, the peak of the conductivity shifts towards higher frequencies, while the conductivity in the low-frequency range decreases rapidly. This results in a reduction of radiation at low frequencies. Although the conductivity increases at high frequencies, the number of photons is very small, so the radiation peak at high frequencies will also be reduced.

The system gives maximal angular momentum radiation at $0.16eV$ ($\sim 4.41\mu m$ ) when $\alpha=0.10eV$  . Electromagnetic waves in this band can be absorbed by many substances, and are therefore commonly used to detect the properties of molecules, gases, and solid materials. They have a wide range of applications, such as in infrared spectrometers and infrared astronomy.

\begin{figure}
	\includegraphics[width=0.9\columnwidth]{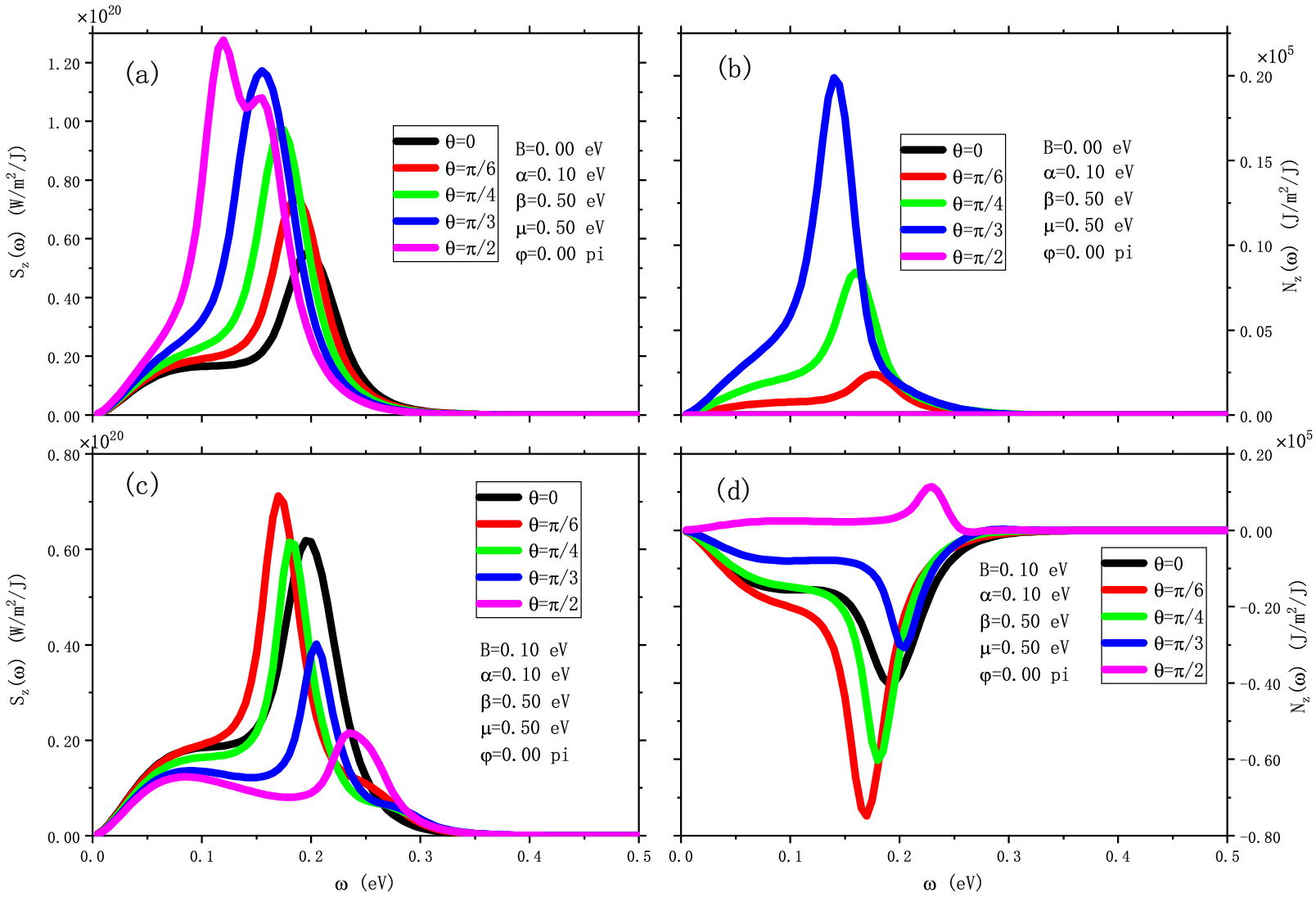}
	\caption{Energy and angular momentum radiation spectrum of altermagnet with different polar angle  . Other parameters are listed in the graph. }
	\label{fig8}
\end{figure}

The modulation of N\'{e}el vector is displayed in Fig.8. The upper figure depicts the energy and angular momentum radiation spectra in the absence of a magnetic field, whereas the lower figure illustrates the case with a magnetic field of $\mu_{B}B=0.10eV$ . In the absence of a magnetic field, the radiation spectrum merely shifts to lower frequencies as the polar angle increases. However, upon applying a magnetic field, the behavior becomes more complex. As the polar angle of the N\'{e}el vector increases, the peak of the energy radiation spectrum initially increases and then decreases, first shifting towards lower frequencies before moving towards higher frequencies.
Regarding the angular momentum radiation spectrum, in the absence of a magnetic field, no angular momentum radiation is generated when N\'{e}el vector is lying in the $xy$-plane or perpendicular to this plane. Between these angles, the peak of the angular momentum radiation spectrum grows with the polar angle, and the peak position moves towards lower frequencies. When a magnetic field is applied, angular momentum radiation is produced for all polar angles, and the direction of the angular momentum radiation is reversed compared to that in the absence of a magnetic field. This indicates that applying a magnetic field allows altermagnets to adjust the magnitude and direction of angular momentum radiation. However, compared to ferromagnetic materials \cite{ref34}, the angular momentum radiation generated by altermagnets is much smaller.

\begin{figure}
	\includegraphics[width=0.9\columnwidth]{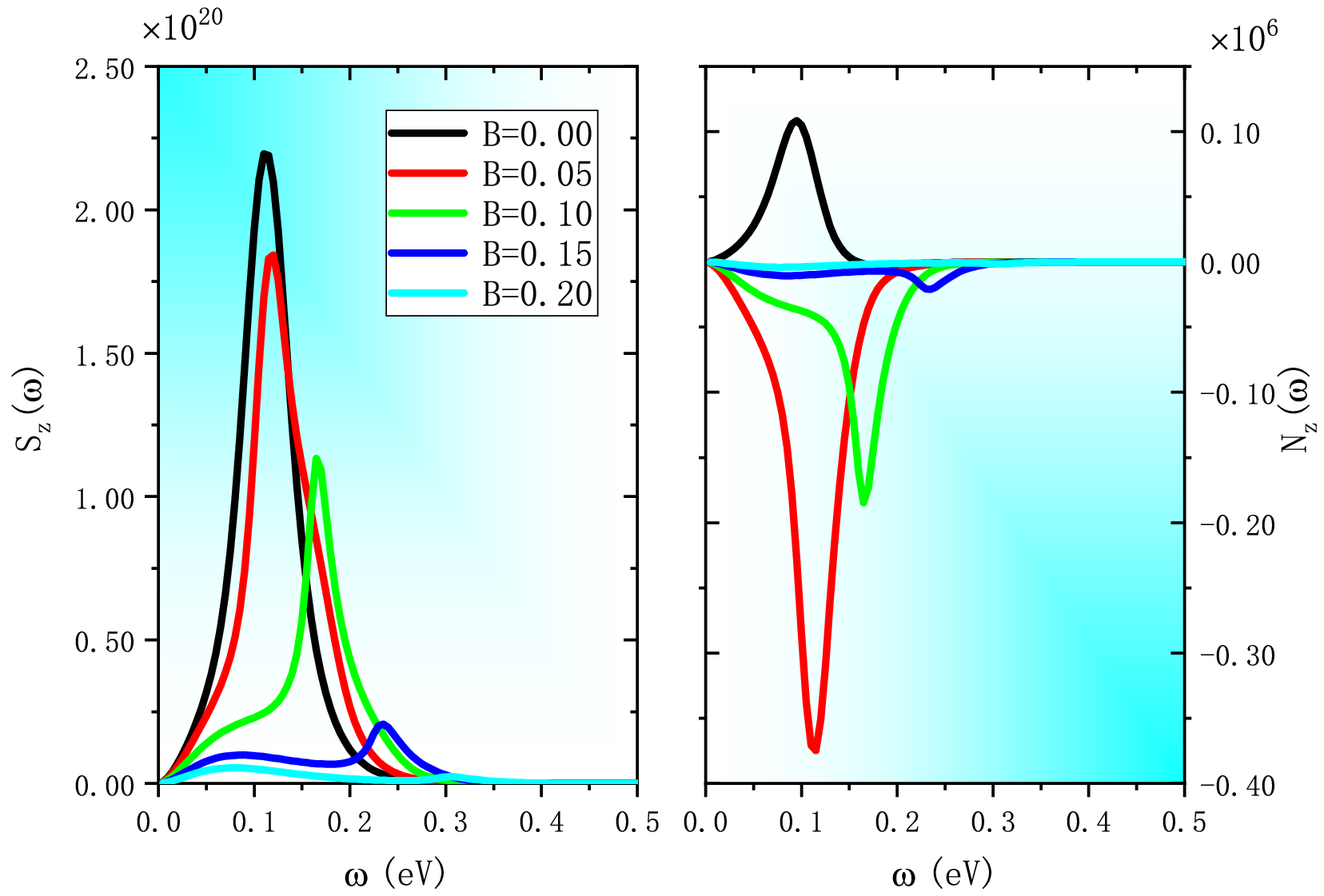}
	\caption{Energy and angular momentum radiation spectrum of altermagnet with different magnetic field. Other parameter are set $\alpha=0.10$, $\beta=0.50$, $\theta=\pi/4$, $\varphi=0.00$, $\mu=0.20$, $T=300K$. }
	\label{fig9}
\end{figure}

 The energy radiation spectrum of altermagnet materials under different external magnetic fields is illustrated in Fig.9. Once a magnetic field is applied, the spectrum peak becomes lower and shifts to a higher frequency. This occurs because the optical conductivity peak moves to a higher frequency with increasing magnetic field, while the number of photons reduced quickly with the increasing of photon energy. Magnetic field brings about significant changes to angular momentum radiation. It alters the direction of angular momentum radiation, in contrast to the scenario without a magnetic field. Even with a small magnetic field, the magnitude of angular momentum radiation is quite enhanced. As the magnetic field increases, the radiation spectrum peak becomes smaller and shifts to a higher frequency.

\subsection{ TOTAL RADIATION}
Integrating radiation spectrum over frequency range will get total radiation. In the following, we investigate how parameters of altermagnet have effects on total energy radiation and angular momentum radiation.

\begin{figure}
	\includegraphics[width=0.9\columnwidth]{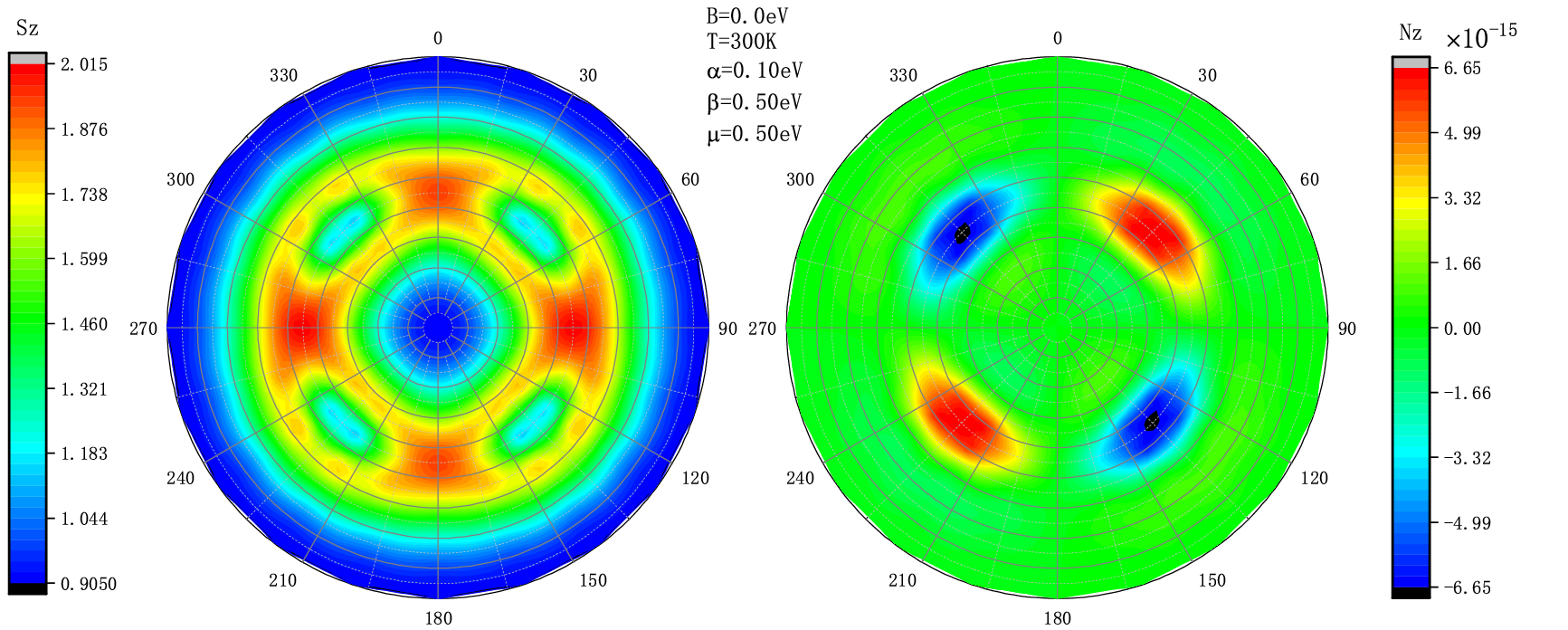}
	\caption{ Energy and angular momentum radiation as a function of $\theta$ and $\varphi$ in the absence of magnetic field. The radial direction is polar angle $\theta$ and the tangential direction is azimuthal angle $\varphi$. }
	\label{fig10}
\end{figure}

In the polar coordinate graph Fig.10, the radial orientation aligns with the polar angle, which spans from $0$ to $\pi$. Conversely, the tangential orientation aligns with the azimuthal angle, extending from $0$ to $2\pi$. Both energy radiation and angular momentum radiation are pronounced when $\theta=\pi/2$, corresponding to N\'{e}el vector in the $xy$-plane. Given the constraints imposed by those parameters, the energy radiation manifests a unique symmetry pattern. Specifically, it shows reflective symmetry about the horizontal and vertical lines, as well as the bisecting lines, with respect to the polar angle. Notably, the energy radiation peaks when the polar angle is near $\pi/2$. Furthermore, it demonstrates quasi-periodic behavior in the azimuthal angle, with a period of $\pi/2$. In the polar coordinate representation, it exhibits a fourfold rotational symmetry ($C_{4}$ symmetry).

The angular momentum radiation exhibits a four-petal shape on the $\theta-\varphi$ diagram,. Two petals have positive values, located in the first and third quadrants, while the other two petals have negative values, situated in the second and fourth quadrants. However, all these values are concentrated around $\theta=\pi/2$. In other regions, the angular momentum radiation values are essentially zero.

From Fig.\ref{fig10} it is concluded that energy and angular momentum are greatly determined by the direction of the N\'{e}el vector in altermagnets. This direction can be effectively regulated through various methods, including spin-orbit torque (SOT), external electric fields, optical control, spin-transfer torque (STT) induced by electric current, and temperature control.

\begin{figure}
	\includegraphics[width=0.9\columnwidth]{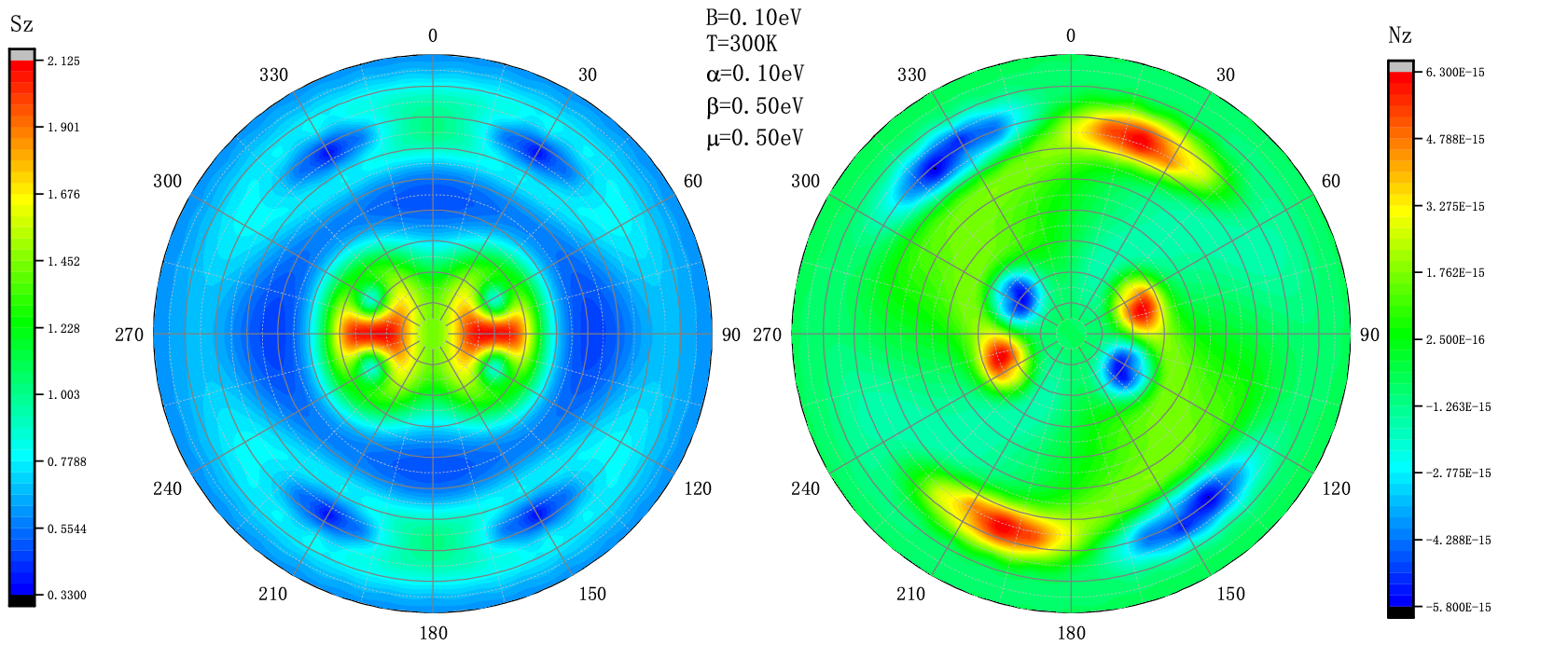}
	\caption{ Energy and angular momentum radiation as a function of $\theta$ and $\varphi$. $B=0.10$. Other parameters are same with Fig.\ref{fig10}.  }
	\label{fig11}
\end{figure}

When a magnetic field is applied, there are significant changes in the energy and angular momentum radiation patterns, as shown in Fig.11. The energy radiation is concentration within a circular boundary where the polar angle $\theta$ , does not exceed $\pi/3$. Within this boundary, two distinct high-intensity patterns emerge, which are associated with the azimuthal angle $\varphi$ being at $\pi/2$ or $3\pi/2$. This is the case that N\'{e}el vector aligns $y$-axis. These patterns signify the regions of maximum radiation. Beyond the circular limit defined by $\theta$ equaling $\pi3$, the energy radiation diminishes significantly. This focused distribution of energy radiation within a defined angular sector is a direct consequence of the magnetic field's effect. The magnetic field has the capacity to modulate the radiation pattern, contingent upon its strength and directional properties. Both energy and angular momentum radiation intensities show $C_{2}$ rotation symmetry, instead of $C_{4}$, due to magnetic field breaking symmetry.

The application of a minor magnetic field to the altermagnet induces substantial alterations in the angular momentum radiation. Across the vast majority of the spatial domain, the angular momentum radiation sustains a value close to zero. Notably, regions of maximal angular momentum radiation, depicted by red and blue patches, emerge when the polar angle $\theta$ aligns with $\pi/4$ or $3\pi/4$ . These patches not only double in number but also reduce in size. This phenomenon underscores the pivotal function of the magnetic field in reconfiguring the distribution of high radiation along the direction of the N\'{e}el vector. The N\'{e}el vector serves as an indicator of antiferromagnetic ordering within the material, highlighting the magnetic field's influence on the material's magnetic structure and radiation characteristics.

The impact of magnetic fields on radiation patterns is profound, facilitating the redistribution and concentration of radiation into particular angular sectors. Moreover, it significantly modifies the distribution and intensity of angular momentum radiation. Understanding these alterations is crucial for elucidating the behavior of altermagnet subjected to magnetic fields. This knowledge is particularly vital in the realm of spintronics and in applications where the magnetic modulation of radiation characteristics plays a pivotal role. The insights gained can lead to advancements in technologies that harness magnetic field interactions with radiative properties, potentially enhancing device performance and functionality in various high-tech applications.

\begin{figure}
	\includegraphics[width=0.9\columnwidth]{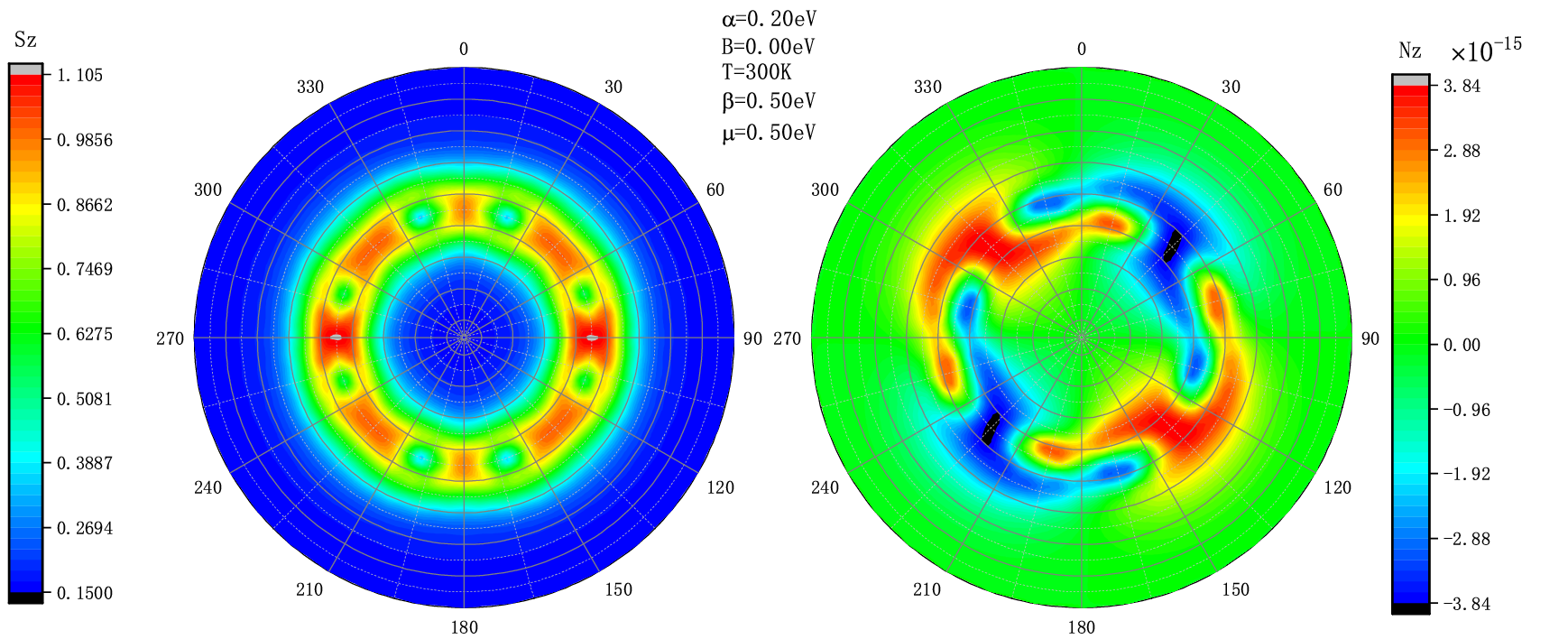}
	\caption{ Radiation of energy and angular momentum radiation as a function of $\theta$ and $\varphi$, with Rashba spin-orbit coupling $\alpha=0.20$. Other parameters are same with Fig.\ref{fig10}.  }
	\label{fig12}
\end{figure}

Fig.\ref{fig12} illustrates the effects of spin-orbit coupling on energy and angular momentum radiation as functions of angles $\theta$ and $\varphi$. The portrayal of energy radiation as a slender band signifies that effective radiation primarily transpires at an angle $\theta$ approximating $\pi/2$ . This suggests a preferential radiation when N\'{e}el vector is perpendicular to $z$-axis. Regarding the azimuthal angle $\varphi$, the energy radiation distribution is non-uniform. The presence of radiation in the stippled zones implies a reduced activity in energy emission for these sectors. Both the inner and outer extents exhibit scant radiation, underscoring their minor role in the aggregate radiation output. This pattern indicates that the principal radiation contributions are localized, with the bulk of the energy being radiated in a specific angular band rather than being diffusely distributed across all possible angles.

The pattern of angular momentum radiation presents a fascinating interwoven structure, characterized by alternating regions of positive and negative angular momentum. The broad areas exhibiting high levels of angular momentum radiation indicate that in the presence of Rashba spin-orbit coupling, altermagnets generate angular momentum radiation easily. There is a notable generation of angular momentum radiation across an expansive range of $\theta$ and $\varphi$ values. This suggests that variations of spin-orbit interaction significantly influence the distribution and intensity of angular momentum radiation.

Both energy radiation and angular momentum radiation display periodic variation with respect to the azimuthal angle $\varphi$. This periodicity implies a structured and repetitive pattern in the radiation characteristics as the azimuthal angle varies, which could be critical for understanding the underlying physical processes and potential applications in fields such as quantum information and condensed matter physics.

The sensitivity of both energy radiation and angular momentum radiation to altermagnet interaction direction, and the spin-orbit coupling strength in the figure with a specified value of $\alpha=0.20 eV$ - underscores the dynamic nature of these radiation patterns. Minor alterations in these critical parameters can precipitate substantial shifts in the observed radiation configurations, reflecting the intricate interplay between quantum mechanical properties and radiative behaviors.

This detailed scrutiny of the figure not only elucidates the principal attributes of the radiation patterns but also contextualizes their ramifications within the framework of spin-orbit coupling phenomena. By accentuating the pivotal role of  $\theta$ and $\varphi$ , the altermagnet interaction, and the spin-orbit coupling strength, the analysis furnishes a structured comprehension that is instrumental for deciphering the complex dependencies inherent in these quantum radiative processes. Such insights are invaluable for advancing theoretical models and experimental designs in the investigation of spin-orbit coupling effects on radiation characteristics, potentially guiding innovations in quantum technologies and material science applications.

\begin{figure}
	\includegraphics[width=0.9\columnwidth]{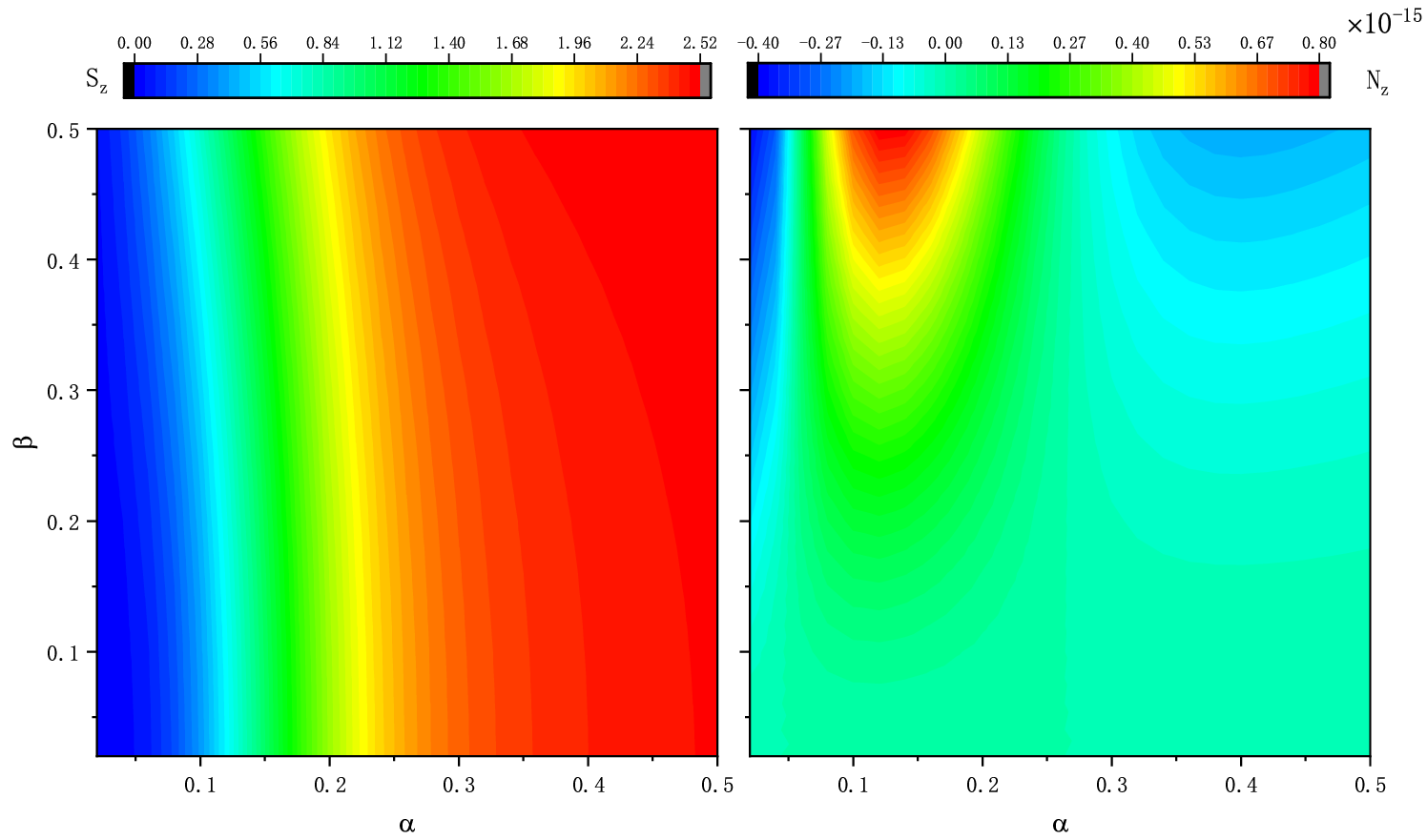}
	\caption{ Energy radiation and angular momentum radiation changes with RSOC strength $\alpha$ and altermagnet interaction strength $\beta$. Other parameters are set $\theta=\pi/4$, $\varphi=\pi/4$, $B=0.0$, $T=300K$, $\mu=0.0$. }
	\label{fig13}
\end{figure}

The modulation of energy radiation in response to variations of spin-orbit coupling $\alpha$ and altermagnet interaction strength $\beta$ is displayed in Fig.\ref{fig13}. Notably, the energy radiation exhibits minimal sensitivity to changes in $\beta$, suggesting a relatively stable radiation profile with respect to altermagnet interaction. In stark contrast, energy radiation displays a pronounced increase with the escalation of $\alpha$, particularly within the range of $0.1eV$ to $0.3eV$. During this interval, the energy radiation escalates swiftly, indicating a heightened responsiveness of the radiation output to alterations in  $\alpha$ within this specific domain. Beyond an $\alpha$ value of $0.3eV$, the energy radiation sustains a comparatively elevated level but demonstrates a plateau, with negligible additional augmentation. The peak emissivity rate recorded is $0.55\%$, a figure markedly inferior to that observed in graphene under analogous thermal conditions \cite{Zhang2022Microscopic}. This disparity can be attributed to the significantly reduced conductivity of the altermagnet material, which acts as a limiting factor in the radiation efficiency, thereby constraining the emissivity rate to a substantially lower threshold than that achievable with materials like graphene that possess superior conductive properties.

The graphical representation on the right panel delineates the variation of angular momentum radiation as a function of spin-orbit coupling and altermagnet interaction. When altermagnet interaction is quite small, angular momentum radiation approximates zero and remains invariant with fluctuations of spin-orbit coupling, indicating a decoupling of these parameters within this regime. However, as altermagnet interaction strength ascends, angular momentum radiation undergoes pronounced oscillations, revealing a heightened susceptibility to changes in altermagnet interaction. Notably, when altermagnet interaction attains a magnitude of $0.5eV$, the angular momentum radiation traverses from a maximal negative value to a maximal positive value, showcasing an extensive dynamic range in response to altermagnet interaction modulation. This behavior contrasts with the energy radiation, which is more substantially influenced by variations in substrate spin-orbit coupling $\alpha$.
Compared with ferromagnets, altermagnets produce almost same angular momentum but with half of energy radiation.

The dichotomy in the responsiveness of energy and angular momentum radiation to the parameters $\alpha$ and $\beta$ is evident. Energy radiation is predominantly modulated by RSOC, with significant alterations in its magnitude corresponding to changes in this parameter. In contrast distinction, angular momentum radiation exhibits a greater volatility to perturbations in altermagnet interaction, with its values undergoing substantial and erratic fluctuations as altermagnet interaction strength varies, particularly at higher values of $\beta$ . This differential sensitivity underscores the distinct roles of spin-orbit coupling and altermagnet interaction in governing the radiative properties of the system under study.

\begin{figure}
	\includegraphics[width=0.9\columnwidth]{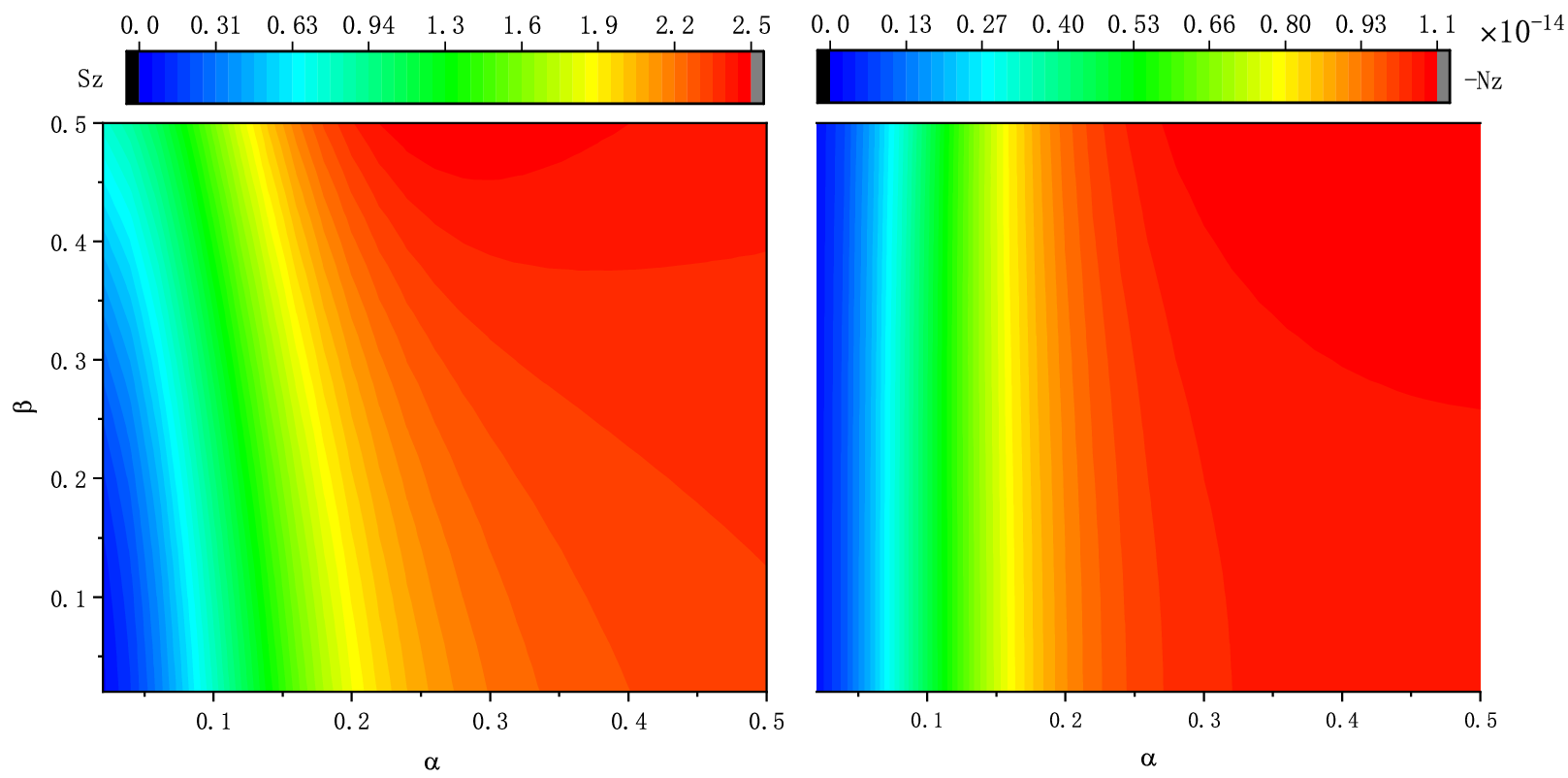}
	\caption{ Energy radiation and angular momentum radiation changes with RSOC strength $\alpha$ and altermagnet interaction strength $\beta$. Other parameters are set $\theta=\pi/4$, $\varphi=\pi/4$, $B=0.05$, $T=300K$, $\mu=0.0$. }
	\label{fig14}
\end{figure}

Figure \ref{fig14} shows the energy radiation and angular momentum radiation properties of antiferromagnets in presence of a weak magnetic field. When a small magnetic field is applied to altermagnet materials, dependence of energy radiation on $\alpha$ and $\beta$ does not change much. Although the energy radiation increases with the increase of $\alpha$ and $\beta$, the energy radiation value is almost the same as that in the absence of a magnetic field.

Instead, angular momentum radiation of altermagnet material is enhanced significantly. With small magnetic field, altermagnets produce pronounced angular momentum radiation. Regardless of the values of $\alpha$ and $\beta$, angular momentum radiation becomes negative, indicating that the direction of the radiated angular momentum is different from that in the absence of a magnetic field. In contrast to the case without a magnetic field, where angular momentum radiation is almost solely dependent on altermagnet interaction, the presence of a magnetic field results in angular momentum radiation that is independent of altermagnet interaction and instead varies with spin-orbit coupling . When spin-orbit coupling is out of consideration, angular momentum radiation is always zero, implying the cancelling between spin-orbit splitting and magnetic field at special  N\'{e}el vector direction. However, as soon as spin-orbit coupling is switched on, angular momentum radiation increases significantly with $\alpha$ and saturates rapidly when $\alpha$  exceeds a medium value $0.2eV$. Both spin-orbit coupling and magnetic fields can induce spin-splitting and affect transverse conductivity and angular momentum, their underlying mechanisms and specific effects differ based on the nature of the interaction (relativistic vs. classical magnetic interaction).

\section { CONCLUSION}
We studied energy and angular momentum radiation of altermagnets in the presence of Rashba spin-orbit coupling and external magnetic field. Energy radiation is notably sensitive to changes in RSOC strength $\alpha$, with significant increases observed particularly between the values of $0.1eV$ and $0.3eV$. Beyond $0.3eV$, the radiation plateaus indicate a saturation effect. The maximum emission rate achieved is$\sim0.55\%$, significantly lower than that of graphene, primarily due to the material's low and narrow peak conductivity. Altermagnets produce angular momentum radiation although the net magnetic moments is zero. These characteristics arise from the breaking of Parity-Time (PT) symmetry in their crystal structures. Due to crystallographic changes that break PT symmetry, non-zero local magnetic moments are generated through spin-orbit coupling under zero external perturbation or PT-symmetric external perturbation. This unique magnetism and electronic structure enable altermagnet metals to produce angular momentum radiation under certain conditions, which is not common in graphene, silicene, or semi-Dirac materials \cite{ref5,ref34,ref35}. Angular momentum radiation shows a heightened sensitivity to altermagnet interaction, contrasting with the energy radiation's primary response to RSOC. N\'{e}el vector direction also has significant influence on both energy and angular momentum radiation. Small changes in these parameters can lead to substantial alterations in the observed radiation, indicating a complex interplay between quantum mechanical properties and radiative behaviors.

The findings underscore the importance of understanding and controlling these parameters for applications in spintronics and other quantum technologies. The ability to manipulate radiation patterns through adjustments in spin-orbit coupling, altermagnet interaction strength and N\'{e}el vector direction could lead to the development of novel devices with tailored radiation characteristics. Further investigation into the mechanisms underlying these radiation process and their sensitivities is warranted. Exploring how external factors such as temperature, magnetic field strength, and material composition influence these parameters could provide additional insights and control over the radiation properties. Altermagnets indeed have many advantages, such as low heat radiation, while being able to generate angular momentum radiation. They are expected to become the ideal candidate materials for next-generation spintronic devices.

\section{Acknowledgements}
The work is partially supported by Basic Research fund of Nanjing University of Aeronautics and Astronautics with Grant No.56XCA2405003.

\bibliographystyle{apsrev4-1}
\bibliography{ALTERMAGNETS_v3}
\end{document}